\documentstyle[12pt]{article}
\begin{document}

\title{Massless
black hole pairs in string theory\\}

\author{\\R. Emparan\\
{\small\sl Department of Physics}\\
{\small\sl University of California}\\
{\small\sl Santa Barbara, CA 93106}\\
{\small\it emparan@cosmic1.physics.ucsb.edu}}

\date{}

\maketitle

\begin{abstract}
We analyze the zero mass black holes that arise as
classical solutions to low energy heterotic string theory. Though these
solutions contain naked singularities, it has been conjectured that
they should be allowed in the theory.
We find a solution describing a pair of oppositely charged
massless black holes in uniformly
accelerated motion {\it under no external force}. By analytically
continuing the solution to Euclidean time, we find an instanton
mediating the pair production of these objects in Minkowski space. 
We analyze the creation
rate, and discuss some consequences of the result.
\end{abstract}

\vfil

\begin{flushleft}
UCSBTH-96-17 \\
hep-th/9607102
\end{flushleft}

\newpage\pagestyle{plain}

Among the many solitonic excitations that have been recently shown to
play a crucial role in the non-perturbative aspects of string theory,
some of the most remarkable ones are zero mass extreme black holes.
They first attracted a great deal of attention when Strominger argued
that they were needed
in order to understand the conifold singularities that
appear in the low energy
Lagrangian describing the moduli space of Calabi-Yau vacua of (type II)
string theory \cite{andy}. The origin of these singularities can be
explained as coming from the integration of quantum loops of black holes
that become massless precisely at the conifold point. One is then
led to include the extremal black holes in the theory in
the same footing as elementary particles.

On the other hand, several classical pointlike solutions of heterotic
string theory have been recently found which have vanishing ADM mass,
as measured at infinity, but
nevertheless carry electric and magnetic charges, and follow timelike,
instead of lightlike, trajectories \cite{behrndt,kallosh,cvetyoum,ortin}.
These solutions are extremal and saturate Bogomolnyi bounds.
One could be tempted
to relate them to the massless black holes described above. However, we
should keep in mind that Strominger's massless holes arise only when
the low energy
action becomes singular and the semiclassical description breaks down,
whereas the classical massless holes
are solutions of a well-defined, non-singular theory.
In the following, we
will focus exclusively on the latter, classical massless holes.

A problem posed by these solutions is the presence of
naked singularities at the position of the hole. These are certainly
required to account for the masslessness of these objects, and are
related also to
other striking features.
These putative massless black holes are peculiarly distinct
from the more ordinary massive black holes.
Their gravitational mass is zero only when measured at
infinity. At finite distances, they exert a repulsive force on massive
test particles \cite{kallosh}, and in this sense they hardly deserve to
be described as `black'. Another puzzling feature of these
massless black holes concerns their kinematical properties.
The string theory massless hole solutions found thus far describe
them at rest.
It is not clear how this state can be changed; for example, Newton's
law does not provide an answer to whether a zero mass hole accelerates
when an external force field is
applied to it.
How can such objects be created? They do not seem to be the extremal
limit of any non-extreme black hole. Therefore, it is unlikely that
they can appear as the last stage in the evaporation of a larger,
non-extremal black hole.
Due to their masslessness and repulsive interactions, it is also
difficult to imagine how they can be formed in gravitational
collapse. There remains the possibility that they are created in pairs.
This mechanism could be analyzed by applying the
Euclidean instanton techniques used in recent years to study
black hole pair creation.

In the present work we address two of these issues, namely, the accelerated
motion of the massless holes, and their spontaneous pair production. The
approach to both problems will be
based on a solution describing a pair of massless black holes
following uniformly accelerated (timelike) trajectories.
A striking property
of the solution is that {\it no external force is needed to accelerate the
holes}. The Lorentzian solution describing this accelerated motion can be
analytically continued to Euclidean
time and then used to construct an instanton that describes the decay of the
Minkowski vacuum by spontaneous
formation of pairs of massless black holes\footnote{Massless black
holes in uniform acceleration have been
also considered in \cite{rasheed} in the context of theories with
imaginary electric charge. However, these massless black holes are not
extremal, and no instanton describing their pair creation could be
constructed. It has also been
argued in \cite{us} that pair creation of (Kaluza-Klein)
massless monopoles can take place in a similar context, namely, in
theories with internal time directions.
In these papers the massless objects
are more intended to show the pathological features of the theories in
which they appear.}.

We find it convenient to present the massless hole solutions from the
same point of view as in
\cite{ortin}, i.e., as bound states of positive and negative mass objects.
The starting point in this interpretation
is a theory with four $U(1)$ gauge fields and
three scalars, described by the action \cite{rahmfeld}
\begin{eqnarray}
\label{hetaction}
I&=&{1\over 16\pi G}\int d^4x\sqrt{-g}\biggl\{ R-{1\over 2}\left[
(\partial\eta)^2 + (\partial\sigma)^2 +(\partial\rho)^2\right]\nonumber\\
&-&{e^{-\eta}\over 4}\left[ e^{-\sigma-\rho}F_1^2 + e^{-\sigma+\rho}F_2^2
+e^{\sigma+\rho}F_3^2 +e^{\sigma-\rho}F_4^2\right]\biggr\},
\end{eqnarray}
which arises as a truncation of the low energy effective action of
heterotic string theory compactified on a six-torus. The equations
of motion of this theory are solved by
\begin{eqnarray}
\label{rahmsol}
ds^2 & = & -(\Delta_1 \Delta_2  \Delta_3  \Delta_4)^{-1}dt^2 +
 (\Delta_1 \Delta_2  \Delta_3  \Delta_4) d{\bf x}^2,\nonumber\\
e^{-\eta}&=&\frac{\Delta_1\Delta_3}{\Delta_2\Delta_4}, \, \,
e^{-\sigma}=\frac{\Delta_1\Delta_4}{\Delta_2\Delta_3}, \, \,
e^{-\rho}=\frac{\Delta_1\Delta_2}{\Delta_3\Delta_4},\\
F_{1/3 \ tj} &=&\partial_j \Delta_{1/3}^{-2}, \quad
\tilde{F}_{2/4 \ tj} =\partial_j\Delta_{2/4}^{-2},
\nonumber\\
\Delta_i &=& \left(1+{q_i\over r}\right)^{1\over 2}\quad i=1,\dots, 4,
\nonumber
\end{eqnarray}
where $\tilde{F}_{2/4}=e^{-\eta\pm (-\sigma+\rho)}*F_{2/4}$, $*F$
being the Hodge dual of $F$.
This notable solution, first found in \cite{cvetyoum2} (see also 
\cite{cvetsey}) and later in
\cite{rahmfeld}, suggests that black hole configurations can be
thought of as composites of four basic building blocks, each charged
under a different Maxwell field \cite{rahmfeld}. 
In particular, the black hole
solutions of single scalar-Maxwell
theories with $a=\sqrt{3},1,1/\sqrt{3},0$ can be obtained
as bound states of one, two, three or four Kaluza-Klein black holes.
Notice that the constituents do not exert any net force on each other
and therefore can be freely superimposed.
The ADM mass of these configurations is
\begin{equation}
\label{mass}
m = {1\over 4}\sum_{i=1}^4 q_i.
\end{equation}
Therefore, it was proposed in \cite{ortin} that
massless solutions could be obtained by allowing for negative
values of the parameters $q_i$.
For example, by
choosing $q_1=-q_3=q$, $q_2=q_4=0$,
we obtain a bound state of  negative mass/positive
mass holes with zero total mass. One should keep in mind that,
as stressed in \cite{ortin},
there is no rigorous way to define the mass of the constituents, even more
in this case since negative mass objects are not expected
to exist in isolation.

We have found that, remarkably, the bound state interpretation can
also be extended to a more complicated class of geometries,
the C-type metrics describing accelerating black holes. The solution is
\begin{eqnarray}
\label{cmetric}
ds^2 &=& {1\over A^2(x-y)^2}\biggl[{\cal F}(x)
\left( {1-y^2\over {\cal F}(y)}
dt^2-
{{\cal F}(y)\over 1-y^2}dy^2\right)\nonumber\\
 &+& {\cal F}(y)\left({{\cal F}(x)\over 1-x^2}dx^2 -
{1-x^2\over {\cal F}(x)}d\varphi^2\right)\biggr],\\
{\cal F}(\xi) &=& \prod_{i=1}^4 f_i(\xi),
\qquad f_i(\xi)= (1-q_i A\xi)^{1\over 2},\nonumber\\
e^{-\eta}&=&\frac{f_1(y) f_3(y)}{f_2(y) f_4(y)}
\frac{f_2(x) f_4(x)}{f_1(x) f_3(x)}, \, \,
e^{-\sigma}=\frac{f_1(y) f_4(y)}{f_2(y) f_3(y)}
\frac{f_2(x) f_3(x)}{f_1(x) f_4(x)}, \nonumber\\
e^{-\rho} &=& \frac{f_1(y) f_2(y)}{f_3(y) f_4(y)}
\frac{f_3(x) f_4(x)}{f_1(x) f_2(x)},\nonumber\\
F_{1/3 \ ty} &=&\frac{q_{1/3}\sqrt{1-q_{1/3}^2A^2}}{f_{1/3}(y)^4},
\nonumber\\
\tilde{F}_{2/4 \ ty} &=&
\frac{q_{2/4}\sqrt{1-q_{2/4}^2A^2}}{f_{2/4}(y)^4}.
\nonumber
\end{eqnarray}
This solution contains as particular cases the dilaton and
$U(1)^2$
extremal C-metrics found
in \cite{dilatonc,ross} for $a=\sqrt{3},1,1/\sqrt{3},0$
(some rewriting of the parameters and
coordinates may be needed to bring the solutions in this form).
One can also add background Melvin fields corresponding
to each of the four $U(1)$ gauge fields, which also take the same
factorized form, and thus obtain Ernst-type metrics.
A detailed analysis of the massive solutions in background fields
will be given elsewhere. For the moment we will only focus
on the solution (\ref{cmetric}) with the purpose of studying the
massless case.

The interpretation of the solution (\ref{cmetric})
in terms of two objects
accelerating apart can be found as usual \cite{kinnersley}, so we shall
be brief. We restrict
the parameters to satisfy $|q_i A|<1$ $\forall i=1,\dots, 4$.
If $|q_i A|>1$ $\forall i=1,\dots, 4$, then
we find an isometry, up to a global sign, 
to the previous case by taking $y\rightarrow y^{-1}$,
$x\rightarrow x^{-1}$, $q_i A\rightarrow (q_i A)^{-1}$.
The static solution (\ref{rahmsol})
can be recovered by setting $y\rightarrow
-1/(rA)$, $t\rightarrow At$, and then taking the limit $A\rightarrow 0$.
This allows us to say that the metric contains objects in one-to-one
correspondence with the solutions of (\ref{rahmsol}).
On the other hand, sending $q_i\rightarrow 0$
yields Rindler (flat) space in
non-standard coordinates.
If there are negative roots $y=(q_i A)^{-1}<-1$, they
correspond to the singularities at the origin
of negative mass constituents, whereas
$y=-\infty$ is interpreted as the position of
the horizon of a positive mass constituent. Also,
$y=-1$ is a Rindler horizon. The variables
are restricted to $-1\leq x\leq 1$, $y_o < y <x$, where
$y_o=\max_i \{ -q_i A, 0\}$. Asymptotic infinity is at $x=y$;
in particular, $x=y=-1$ is spacelike infinity.
The parameter $A$ roughly measures the
acceleration of the black holes,
but also their separation when they become
closest, which is given by (twice) $1/A$.
This solution asymptotes to
flat space (with,
possibly, a conical singularity, see below).

The angular part $(x,\varphi)$ of the metric, with
$\varphi$ identified with period $\Delta\varphi$,
has the topology
of a two-sphere. In general, one can not choose $\Delta\varphi$
so as to eliminate conical singularities at both poles $x=\pm 1$. These
conical singularities reflect the need of forces to accelerate the
black holes,
and are usually cancelled by, e.g., adding background fields.
However, let us examine this issue closer for the metric (\ref{cmetric}),
which does not contain such fields.
Regularity at both poles can only be achieved if
\begin{equation}
\Delta\varphi = 2\pi {\cal F}(1) =2\pi{\cal F}(-1).
\end{equation}
For arbitrary $A$ the equation ${\cal F}(1) ={\cal F}(-1)$ requires
\begin{equation}
\label{noforce}
q_1+q_2+q_3+q_4 =0
\end{equation}
and
\begin{equation}
\label{noforce2}
q_1 q_2 q_3 +q_1 q_3 q_4 +q_2 q_3 q_4 +q_1 q_2 q_4=0.
\end{equation}
But (\ref{noforce}) is just the zero mass condition for the black holes!
Though the mass of the black hole is not well defined in the
non-spherical metric (\ref{cmetric}), it should be clear that the
object that is accelerating is in direct correspondence to the
static, massless hole.
Therefore, by choosing the parameters to satisfy these equations,
we succeed in finding a solution
describing two massless
black holes moving with uniform acceleration in an asymptotically
Minkowskian space. No external force is acting upon the black holes.

The meaning of (\ref{noforce2}) is less clear, but it is probably
required to cancel
the forces that appear between black holes when they are in motion. In any
case, it is satisfied by the simplest and most interesting 
massless holes.

The parameters $q_i$ are not exactly equal to the physical charges
of the black hole. Rather, they approximate them in the limit
$A\rightarrow 0$. The physical charges ${\hat q}_i$ are found by
integrating the (dual) field strengths on a sphere surrounding the black
hole. One finds then
\begin{equation}
{\hat q}_i=q_i{{\cal F}(1)\over \sqrt{1-q_i^2 A^2}}.
\end{equation}

An aspect worth mentioning is that the metric (\ref{cmetric})
does not become spherical
as the position of the hole is approached. The hole
is slightly elongated along the axis of acceleration. A similar effect was
also found in \cite{ross}. This is in contrast to
\cite{dilatonc}, where the geometry close to an
extremal dilaton black hole in accelerated motion was
found to approach exactly that of the static hole.

The bound state interpretation
suggests that a small perturbation of the static massless holes
could set them in runaway motion, in which the negative mass constituent
chases the positive mass one (this is related to the old problem of
``Bondi dipoles'' in General Relativity, see e.g. \cite{israel}).
Our C-metric precisely describes such motion
(though we should keep in mind that the solution
actually describes {\it a pair} of massless holes).
The elongation of the massless hole
can presumably be seen as the separation between
the distributions of negative and positive massess needed
to maintain the continued accelerated motion of the bound state. 

It might seem puzzling that the interaction between the two
massless holes, which one would expect to be attractive since they
form a particle-antiparticle pair, does not prevent the
pair from separating away. The holes are 
accelerated and distorted, and the forces between them are rather 
complicated. It is clear that these forces should be
more and more negligible as we take $A\rightarrow 0$ (and therefore
$1/A\rightarrow \infty$). The fact that, in the absence of
external forces, the holes accelerate
apart instead of collapsing and annihilating each other, would suggest 
that their net effective interaction could in fact be 
repulsive \footnote{This might be related to the appearance
of negative Euclidean action that we find below.}. Notice, however, 
that it is
not possible to readily resort to the 
repulsive forces found in \cite{kallosh}, since the
massless holes cannot be treated as test particles. In any case, the
solution we have found certainly incorporates all these interactions, 
and the absence of external forces appears neatly in the final result.

Now, the solution (\ref{cmetric}) can be continued to Euclidean
time by taking
$\tau=i t$. Regularity at the acceleration horizon at $y=-1$ is achieved
by taking the period of $\tau$ to be
\begin{equation}
\beta= 2\pi{\cal F}(-1).
\end{equation} 
This Euclidean metric describes a massless
black hole loop.
By slicing in half the instanton, and
glueing it to the time-symmetric section of the Lorentzian solution,
we obtain a description of the spontaneous creation of a pair of massless
black holes out of the Minkowski vacuum.

What is the creation rate for this process? This is usually taken
to be given, to
leading order, by $e^{-I}$, where $I$ is the classical action of the
Euclidean instanton. Following \cite{hawking}, we decompose the action in
the following terms:
\begin{equation}
\label{eucact}
I=\beta H -{1\over 4}(A_{bh}+\Delta A)
\end{equation}
with $H$ the Hamiltonian, $A_{bh}$ the area of the black hole horizon,
and $\Delta A$ the difference in the area of acceleration horizons.
The action is calculated with reference to a background which in the
present case will be flat space.

The area term $A_{bh}/4$ is associated, as usual, to the entropy of
the black hole, and
it has been argued that this term must be present, in string theory,
for non-extreme as well as for extreme black holes \cite{gary}.
A potentially troublesome issue in the case at hand is the curvature
singularity at the position of the hole. There is no horizon area
at that point. However, we are considering the possibility
that massless hole singularities could be allowed in the theory.
One could, perhaps, expect the geometry
near the horizon to be modified
where the curvature reaches the string scale and string
effects can be expected to come into play.
It has been argued that,
for zero area black holes in
string theory, the entropy of the black hole can be found by drawing a
surface at that radius \cite{sen}. We would then find a term $S_{bh}$
enhancing the pair creation rate.
Far from the horizon, string theory corrections should in any case
be negligible,
so that $H$ and $\Delta A$ can be computed
in the classical, low energy geometry.

The Hamiltonian in (\ref{eucact}) consists of two terms,
a bulk term that vanishes on exact solutions, and a surface term.
Since the spacetime is non-compact, we
must introduce a boundary near infinity; care must be taken so that
the boundaries of the C-metric and the flat space background are
identical. When this is done,
a detailed but straightforward calculation
shows that the surface term in the Hamiltonian vanishes, just as in
all black hole pair creation processes analyzed thus far.

We are therefore left with
\begin{equation}
I=-{1\over 4}\Delta A -S_{bh}.
\end{equation}
$\Delta A$ should give the dominant contribution to the pair
creation rate. In fact, it usually
gives the total contribution to the pair creation rate of extreme 
black holes
\cite{hawking}. Again, to calculate it we must carefully
perform the subtraction of the
areas. By introducing a boundary in the C-metric at
$y=-1$, $x=-1+\epsilon_c$ and evaluating the quantities only up to
second non-trivial order in $\epsilon_c$,
the length of this boundary is
\begin{equation}
l_c=\int d\varphi \sqrt{g_{\varphi\varphi}} =
2\pi \rho_c\left(1-{\sum_i {\hat q}_i^2\over 2\rho_c^2} -
{{\cal F}(1)^2 \over
2 A^2 \rho_c^2}\right),
\end{equation}
where the no-force conditions (\ref{noforce},\ref{noforce2}) have
been imposed and we have defined $\rho_c\equiv {\cal F}(-1)
\sqrt{2\over A^2\epsilon_c}$.
The boundary in the Minkowski background, written in cylindrical
coordinates $(t,z,\rho,\varphi)$, is at $\rho=\rho_M$, and its length is
$l_M=2\pi\rho_M$. We match the boundaries in the C-metric
and in Minkowski spacetime by requiring $l_M=l_c$.

The area of the acceleration horizon in the C-metric is
\begin{equation}
A_c=\int dx\; d\varphi\sqrt{g_{xx} g_{\varphi\varphi}}|_{y=-1}=
-{\pi{\cal F}(1)^2\over A^2} + \pi\rho_c^2.
\end{equation}
In the background, the acceleration horizon is at $z=0$, and the area
inside a circle $\rho=\rho_M$ is $A_M=\pi \rho_M^2$.
Then we end up with the exact result
\begin{equation}
\label{difarea}
\Delta A=A_c-A_M=\pi \sum_{i=1}^4 {\hat q}_i^2.
\end{equation}

This is a striking result, and not only because of its simplicity.
Notice that $\Delta A>0$, i.e., the area of the acceleration horizon
in the C-metric is
{\it bigger} than in the reference metric. This is just the opposite
of what happens in all the black hole creation processes studied
thus far, where, in fact, this term
yields a negative result and thus the suppression
factor for the pair creation rate. For the massless holes, however,
the leading term in the pair creation rate should be
$\exp ({\pi\over 4}\sum_{i=1}^4 {\hat q}_i^2)$, (times, possibly, a
factor $e^{S_{bh}}$).
Apparently, pair production
of massless black holes is not suppressed,
but, on the contrary, enhanced! \footnote{The reader should be cautioned,
though, that
there is no agreement between different authors as to whether the action
contribution should be weighed as $e^{-I}$ or rather as $e^{-|I|}$.}.
This is certainly disturbing, since it would seem to imply that if these
massless holes were actually possible, the
Minkowski vacuum should be highly unstable against their production
in pairs.

A minor subtlety arises in that, when continuing to Euclidean time, 
the 
electric-type charges $q_1$, $q_3$, should also be continued to imaginary 
values. It has been argued, however, that in these cases a 
charge projection on electric states
must be introduced that restores the expected
electric-magnetic duality of decay rates \cite{hawross}. Therefore, we
still expect the Euclidean exponential pair production factor to be 
given by the same positive definite function 
(\ref{difarea})\footnote{Notice that
it is always possible to have a massless hole constructed out of only
magnetic-type constituents, i.e., $q_2=-q_4$, $q_1=q_3=0$.}.

The result is also peculiar in other respects. Normally,
for extreme massive
black holes, the term $-\Delta A/4$ when expressed in terms of physical
parameters reproduces the usual exponent of the
Schwinger pair production rate
\begin{equation}
I={\pi m^2\over F}(1+ O(F)) ={\pi m\over A}(1 + O(m)),
\end{equation}
where $F$ is the driving force, $F\approx mA$. In the limit of
zero mass, one would expect to find vanishing action. Certainly, this
is very different from the behavior that we have found.
Moreover, creation of
objects in slowly accelerated motion is usually strongly suppressed
since the action grows larger. In contrast, the corresponding term
(\ref{difarea}) goes
smoothly to a finite value as $A\rightarrow 0$.
Thus, within the range $|q_i A|<1$, the massless black holes can be created 
with
arbitrary acceleration. Recalling that $1/A$ measures the separation
between black holes at the moment of their production, we see
that the massless holes can be created arbitrarily far from each other and we
still find a finite, bounded action.

Let us now discuss some potential problems of the result we have found.
The static extremal solutions are protected from quantum
renormalization
effects by supersymmetry. The solution
(\ref{cmetric}) does not presumably preserve any supersymmetry.
However, since we can create the black
holes far apart from each other, they can approximate more and more
the static solutions. Recent experience
suggests that, in some cases, corrections in near 
BPS states can be kept under control
\cite{andygary}. 
We also want to remark that
our analysis does not depend on the bound state interpretation. One
could simply take the solution and not bother about hypothetic
constituents, though the form of the solutions is certainly compelling.

This troublesome unsuppressed pair creation out of the vacuum does not
readily apply to the quantum massless black holes of 
\cite{andy}. The fact that the asymptotically 
Euclidean instanton above
has negative action is clearly due to the naked singularities
of the massless holes. This may cast some doubt on whether these 
classical
singular solutions shoud be allowed in the theory.
As already mentioned, string effects in the region
near the hole can modify the geometry and correct the total
action, but it is not clear how this could change qualitatively the result.
The pair production mechanism described is based on commonly accepted
semiclassical instanton methods. Should one reject their validity
when massless holes are present? For example, it could be objected that,
usually, semiclassical instanton techniques require the action to be 
large (and positive), which is not the present case. But this is more an
objection to the quantitative result than to the qualitative instability
suggested by the solution above. We also remark that the unusual result
$\Delta A>0$ is already present in the Lorentzian regime.
With all these caveats in mind, the conclusion is that the pair creation
process described clearly illustrates that allowing for classical
massless solutions in the theory may be a source of instabilities.

We have benefited from helpful comments from G.~Horowitz, 
A.~Strominger, and T.~Ort{\'\i}n.
This work has been partially supported by a FPI postdoctoral fellowship
from MEC (Spain), and by CICYT AEN-93-1435 and UPV 063.310-EB225/95.

\end{document}